\documentclass{ckm}                 

\def\Journal#1#2#3#4{{#1} {\bf #2}, #3 (#4)}

\def\NIMA{{\em Nucl. Instrum. Methods} A}
\def\NPB{{\em Nucl. Phys.} B}
\def\PLB{{\em Phys. Lett.}  B}
\def\PRL{\em Phys. Rev. Lett.}
\def\PRD{{\em Phys. Rev.} D}
\def\ZPC{{\em Z. Phys.} C}
\def\ANNU{{\em Ann.Rev.Nucl.Part.Sci.}}

\def\be{\begin{equation}}
\def\ee{\end{equation}}
\def\bea{\begin{eqnarray}}
\def\eea{\end{eqnarray}}

\def\ra{\rightarrow}

\def\dkln{D^0 \rightarrow K^-l^+\nu_l}
\def\dkpln{D^+ \rightarrow K^-\pi^+l^+\nu_l}

\def\dkskpln{D^+ \rightarrow \overline{K}^{*0}(K^-\pi^+)l^+\nu_l}
\def\dsphikkln{D_s^+ \rightarrow \phi(K^-K^+)l^+\nu_l}

\def\dkpmn{D^+ \rightarrow K^-\pi^+\mu^+\nu}

\def\dkskpmn{D^+ \rightarrow \overline{K}^{*0}(K^-\pi^+)\mu^+\nu}
\def\dsphikkmn{D_s^+ \rightarrow \phi(K^-K^+)\mu^+\nu}
\def\costv{\cos \theta_V}
\def\dksmn{D^+ \rightarrow \overline{K}^{*0}\mu^+\nu}
\def\dkpp{D^+ \rightarrow K^-\pi^+\pi^+}

\def\dsphikkp{D_s^+ \rightarrow \phi(K^-K^+)\pi^+}
\def\knp{K\,\pi}

\confname{Workshop on the CKM Unitarity Triangle, IPPP Durham, April 2003}

\title{Recent results on charm from E831-FOCUS}

\author{Daniele Pedrini \\
 on behalf of the FOCUS Collaboration\thanks{Collaboration members are
    J.~M.~Link, P.~M.~Yager (UC Davis);
    J.~C.~Anjos, I.~Bediaga C.~G\"obel, J.~Magnin, A.~Massafferri, 
    J.~M.~de~Miranda, I.~M.~Pepe, E.~Polycarpo, A.~C.~dos~Reis (CBPF);
    S.~Carrillo, E.~Casimiro, E.~Cuautle, A.~S\'anchez-Hern\'andez,
    C.~Uribe, F.~V\'azquez (CINVESTAV);
    L.~Agostino, L.~Cinquini, J.~P.~Cumalat, B.~O'Reilly, I.~Segoni,
    M.~Wahl (Colorado Boulder);
    J.~N.~Butler, H.~W.~K.~Cheung, G.~Chiodini, I.~Gaines,
    P.~H.~Garbincius, L.~A.~Garren, E.~Gottschalk, P.~H.~Kasper,
    A.~E.~Kreymer, R.~Kutschke, M.~Wang (FNAL); 
    L.~Benussi, M.~Bertani, S.~Bianco, F.~L.~Fabbri,
    A.~Zallo (INFN Frascati);
    M.~Reyes (Guanajuato); 
    C.~Cawlfield, D.~Y.~Kim, A.~Rahimi, J.~Wiss (Illinois Urbana-Champaign);
    R.~Gardner, A.~Kryemadhi (Indiana Bloomington); 
    Y.~S.~Chung, J.~S.~Kang, B.~R.~Ko, J.~W.~Kwak, 
    K.~B.~Lee (Korea University);
    K.~Cho, H.~Park (Kyungpook);
    G.~Alimonti, S.~Barberis, M.~Boschini, A.~Cerutti,  P.~D'Angelo,
    M.~DiCorato, P.~Dini, L.~Edera, S.~Erba, M.~Giammarchi,
    P.~Inzani, F.~Leveraro, S.~Malvezzi, D.~Menasce, M.~Mezzadri,
    L.~Moroni, D.~Pedrini, C.~Pontoglio, F.~Prelz, M.~Rovere,
    S.~Sala (Milano and INFN Milano);
    T.~F.~Davenport~III (North Carolina  Asheville);
    V.~Arena, G.~Boca, G.~Bonomi, G.~Gianini, G.~Liguori, D.~Lopes~Pegna,
    M.~M.~Merlo, D.~Pantea, S.~P.~Ratti, C.~Riccardi, P.~Vitulo 
     (Pavia and INFN Pavia);
    H.~Hernandez, A.~M.~Lopez, E.~Luiggi, H.~Mendez, A.~Paris, J.~Quinones,
    J.~E.~Ramirez, Y.~Zhang (Puerto Rico  Mayaguez);
    J.~R.~Wilson (South Carolina Columbia);
    T.~Handler, R.~Mitchell (Tennessee Knoxville);
    D.~Engh, M.~Hosack, W.~E.~Johns, M.~Nehring, P.~D.~Sheldon, K.~Stenson,
    E.~W.~Vaandering, M.~Webster (Vanderbilt); 
    M.~Sheaff (Wisconsin Madison)}}

\address{INFN Sezione di Milano, Milano, Italy}

\begin{document}

\begin{abstract}
  E831-FOCUS is a photoproduction experiment which collected data during the 1996/1997 
fixed target run at Fermilab. More than 1 million charm particles have been reconstructed. 
Using this sample we measure the lifetimes of all the weakly decaying singly charmed particles, 
establishing the charm lifetime hierachy. Then we present recent results on semileptonic decays 
of charm mesons, including the new s-wave inteference phenomena in $D^+ \to K^-\pi^+\mu^+\nu$, 
and high statistics branching ratio and form factor measurements.
\end{abstract}
\maketitle


\section{Introduction}

 Investigations of the $K$ and $B$ systems have and will continue to play a central role in our quest to understand flavor 
physics~\cite{Hewett}, but investigations of the charm-quark sector are fundamental too. 
Since charm is the only {\it up-type} quark for which the decay modes can be studied, 
it has a unique role to investigate flavor physics. Charm allows a complementary probe of 
Standard Model beyond to that attainable from the {\it down-type} sector. Here we present recent 
analyses on lifetimes and semileptonic decays.

 The E831-FOCUS spectrometer is an upgraded version of the E687 fixed target
spectrometer~\cite{spectro}, located in the Fermilab proton beam area, which collected 
data during the 1996--97 fixed target run. Electron and positron beams (with
typically $300~\textrm{GeV}$ endpoint energy) obtained from the $800~\textrm{GeV}$ Tevatron
proton beam, produce by means of bremsstrahlung, a photon beam which
interacts with a segmented BeO target. The mean photon energy for triggered
events is $\sim 180~\textrm{GeV}$. Two systems of silicon microvertex
detectors are used to reconstruct vertices: the first system consists of 4 planes
of microstrips interleaved with the experimental target~\cite{WJohns} and the
second system consists of 12 planes of microstrips located downstream of the
target. These detectors provide high resolution in the transverse plane
(approximately $9~\mu\textrm{m}$), allowing the identification and separation of charm
primary (production) and secondary (decay) vertices. More than 1 million charm particles 
have been fully reconstructed.

\section{Charm lifetimes}

\begin{figure*}[ht!]
\hbox to\hsize{\hss
\includegraphics[width=0.5\hsize]{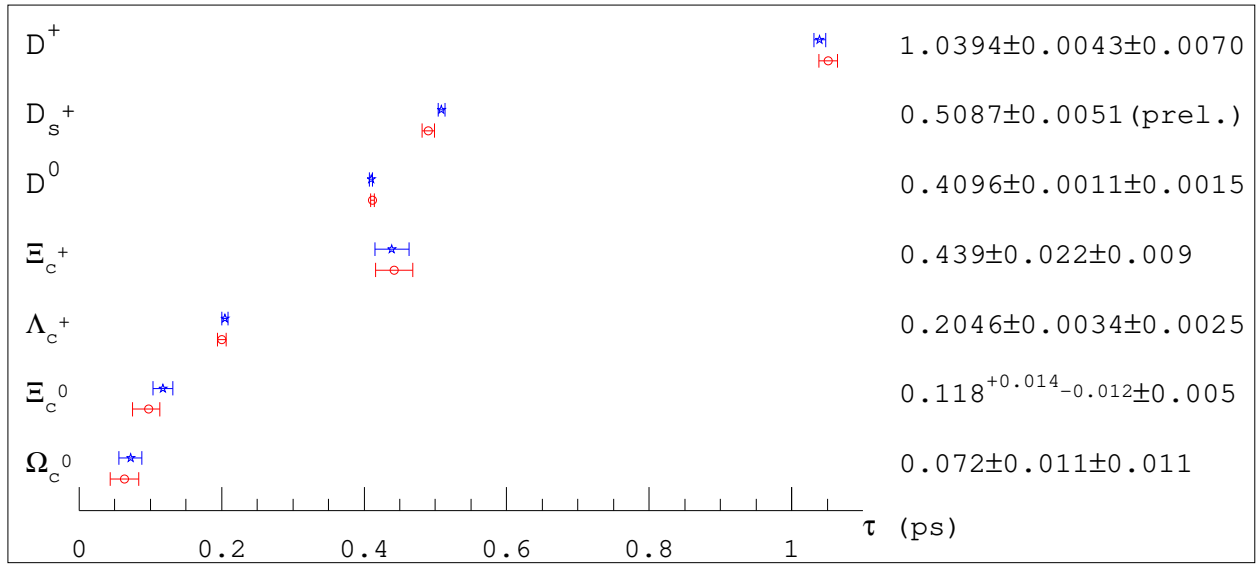}
\hss}
\caption{Charm particle lifetimes, comparison between the FOCUS lifetime 
measurements and the PDG 2002 values. The $\star$ are the FOCUS results reported 
also on the right, while the $\circ$ correspond to the PDG 2002 values. The PDG 2002 
values for $\Xi_c^+$ and $\Lambda_c$ include already our measurements. 
\label{fig:lifetime}}
\end{figure*}

 The determination of lifetimes allows to convert the branching ratios measured by experiments
to partial decay rates predicted by theory. FOCUS is the only experiment (with the 
predecessor experiment E687) to have measured the lifetimes of all the weakly decaying charmed 
particles. This is particulary important when one forms the ratio of lifetimes because 
most of the systematic errors cancel out. In Fig~\ref{fig:lifetime} we show a comparison between 
the PDG 2002~\cite{PDG} values and the FOCUS lifetime measurements (in two cases our results 
are already included in the weighted averages). FOCUS produced new lifetimes results with precision 
better than the previous world average. An accurate measurement of the $D^0$ lifetime for the golden 
decay mode into $K\pi$ is a crucial ingredient to determine the lifetime difference, and consequently 
the parameter $y$ of the $D^0-\overline{D^0}$ mixing. 

 The increasingly precise measurements of the heavy quark lifetimes have
stimulated the further development of theoretical models, like the Heavy Quark Theory~\cite{HQE}, 
which are able to predict successfully the rich pattern of charm hadron lifetimes, that span 
one order of magnitude from the longest lived ($D^+$) to the shortest lived ($\Omega_c^0$).

 For the charm mesons a clear lifetime pattern emerges in agreement with the 
theoretical predictions:

\begin{equation}
\tau(D^0)<\tau(D_s^+)<\tau(D^+)
\end{equation}

 Even the expectations~\cite{Guberina,HQE} for the charm baryon lifetimes reproduce the data, 
which is quite remarkable since, in addition to the exchange diagram, there are constructive 
as well as destructive contributions to the decay rate. The experimental results lead to the 
following baryon lifetime hierarchy :

\begin{equation}
\tau(\Omega_{c}^0)\leq\tau(\Xi_{c}^{0})<\tau(\Lambda_{c}^+)<\tau(\Xi_{c}^{+})
\end{equation}

\section{Semileptonic Decays of Charm Particles}
Traditionally, the semileptonic decays of heavy flavored particles are
accessible to both collider and fixed target experiments. 
The decays have clean and distinguishable signatures, and the Cabbibo-allowed
decay channels like $\dkln$, $\dkskpln$, $\dsphikkln$ 
have large branching ratios. 

Their fully explicit decay rates can be calculated from first principles,
for example, theoretical tools like Feynman diagrams. Involving a lepton
in the final decay stage implies that we do not have to worry about the usual
final state interaction between hadrons.  The possible complications coming
from QCD corrections of the decay process are contained in form factors.
The form factors can be calculated by various methods, Lattice Gauge Theories 
(LGT) and quark models. The angular distributions and invariant masses among the decay
products would determine the form factors ratios while the branching
ratio measurements and information from the CKM matrix would give the absolute
scale for the form factors.  

\subsection{The New S-wave Interference in $\dkpmn$ Decays}
For the last 20 years, people regarded the $\dkpmn$ decays as 100\% $\dkskpmn$
events. The E687 and E691 groups set an upper limit for the possible scalar 
contributions in the  $\dkpln$ decays~\cite{e687,e691}, 
but they could not provide clear evidence of decay paths
other than the dominant P-wave $\dkskpln$ channel.
The situation was changed when the next generation data set from the FOCUS
spectrometer was analyzed to get form factors of the 
$\dkpmn$ decays~\cite{focus_sw}.

After the selection cuts involving vertex confidence levels and particle 
identification requirements, we obtained 31,254 $\dkpmn$ and its 
charge conjugate decays\footnote{In this paper we assume that a decay and
its charge conjugate decay go through the same physical process.}. 
During the form factor analysis,
we checked the angular distribution of Kaon in the $\knp$ rest frame
($\costv)$ and found that it showed a huge forward-backward asymmetry
below the $K^*(892)$ pole mass while almost no asymmetry
above the pole.  Since the $K^*$ is a P-wave, pure
$K^* \ra K\pi$ decays would have shown only a symmetric forward-backward
$\costv$ distribution over the entire $\knp$ invariant mass range. This 
suggests a possible quantum mechanics interference effect.  

   A simple approach to emulate the interference effect is adding a
spin zero amplitude in the matrix elements of the $\dkpmn$ decays. We tried 
a constant amplitude with a phase, $A \exp(i\delta)$, in the place
where the $K^*$ couples to the spin zero component of the $W^+$ particle. 
We made the simplest assumption that the $q^2$ dependence of this anomaly S-wave
coupling would be the same as that of the $K^*$.

\begin{figure}
\hbox to\hsize{\hss
\includegraphics[width=0.5\hsize]{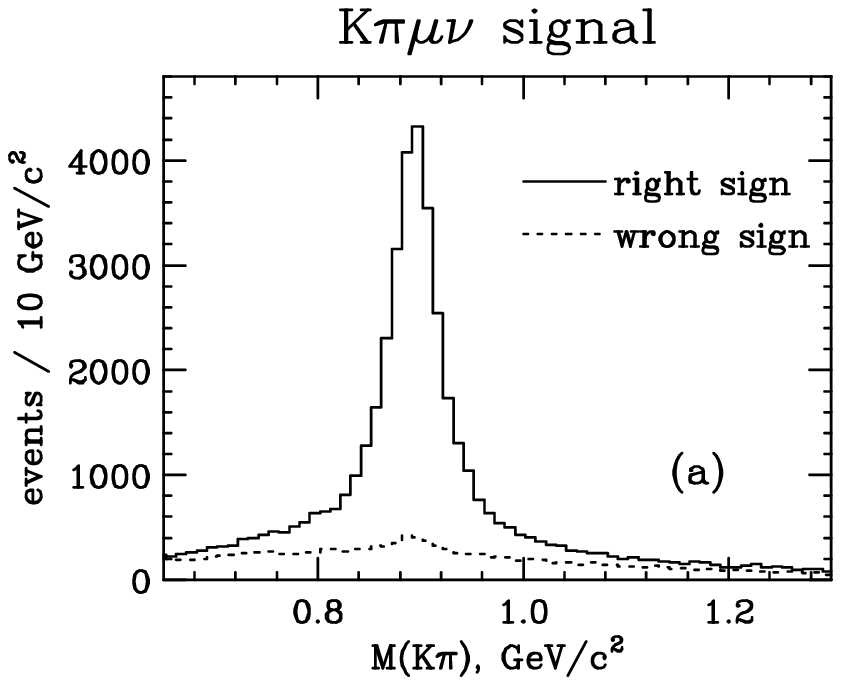}
\hss
\hspace{0.5cm}
\includegraphics[width=0.5\hsize]{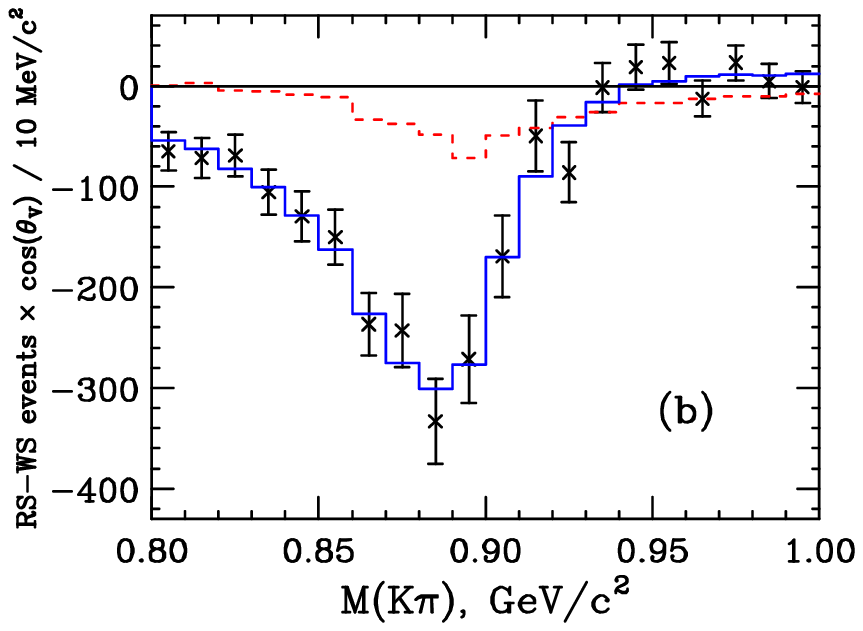}
\hss}
\caption{(a) $\dkpmn$ signal. The wrong-sign-subtracted yield
is 31,254 events. (b) Asymmetry distribution in $\knp$ invariant
mass. The dashed line 
represents Monte Carlo simulation with no interfering s-wave amplitude
while the solid line represents Monte Carlo simulation with an s-wave
amplitude. The points with error bars are the experimental data.
\label{fig:costv_mkpi}}
\end{figure}

   The $\dkpmn$ event is a 4-body decay, which is represented by 5 kinematic
variables, two invariant masses and three angular variables. For each of
these variables, we extracted interference effects by using various weighting
schemes and studied if our measured $A = 0.36$ and $\delta = \pi/4$
are working properly in reproducing the effects for Monte Carlo (MC) 
events~\cite{focus_sw}. 
As shown in Fig.~\ref{fig:costv_mkpi}
where the invariant mass of the $\knp$ particles are weighted by $\costv$,
the interference effect is reproduced with satisfaction. Our measured phase
of $\pi/4$ relative to the $K^*(892)$ is consistent with the one found by LASS
collaboration for isosinglet s-wave around the K* pole from a $\knp$ phase
shift analysis~\cite{lass}. Our data is consistent with a broad 
resonance interpretation as well, but the pole of the resonance would be located
above the $K^*$ pole in absence of any FSI re-phasing.
We tried a $\kappa(800)$ resonance hypothesis. It turned out that 
to produce the interference effect, a 100 degree phase shift is needed
between the $\kappa$ and the $K^*$.

One interesting side effect of the S-wave interference is that it breaks 
the $\chi \leftrightarrow -\chi$ symmetry of the distribution
of the azimuthal angle ($\chi$)
between the $\knp$ and the $W^+$ decay planes in the $D^+$ rest frame. 
The proper definition of $\chi$ requires that it should change sign
between $\dkpmn$ and its charge conjugate decays. Without the proper sign
convention, we would see a false CP violation between the charge conjugate
decays in the $\chi$ distribution.

\subsection{Branching Ratio Measurements}
    We measured the relative branching ratio between $\dksmn$ and $\dkpp$
decays.  With a tighter selection than the one used in the interference
analysis, we selected 11,698 $\dkpmn$ and its charge conjugate decays. With a
selection cut set designed to be similar to the one applied upon the
$\dkpmn$ decays, we obtained 65,421 $\dkpp$ and its charge conjugate decays.
From a MC study, we determined that the pure $\dksmn$ events are 94.5\%
of the selected events. When this correction factor is applied, we
obtained~\cite{focus_br},
\be
 \frac{\Gamma(\dksmn)}{\Gamma(\dkpp)}= 0.602 \pm 0.010\pm 0.021
\ee
When comparing this muon decay channel result with electron decay channel
results from other experiments, a correction factor 1.05 should be applied. Our
number, the only one considered an S-wave interference explicitly, is
1.6 $\sigma$ lower than the recent CLEO II result 
from the electronic decay channel~\cite{cleo} and 2.1 $\sigma$ higher than 
the E691 measurement~\cite{e691_br}.  
Including our result, the new world average of
$\Gamma(K^*l\nu)/\Gamma(K\pi\pi)$ is 0.62 $\pm$ 0.02 each experiment's
statistical and systematic errors were added in quadrature
prior to making the weighted average.

We also measured the relative branching ratio between $\dsphikkmn$
and $\dsphikkp$ decays. Our selection yields 793 $\dsphikkmn$ and its charge
conjugate decays, and 2,192 $\dsphikkp$ and its charge conjugate decays.
The result is~\cite{focus_br} 
\be
 \frac{\Gamma(\dsphikkmn)}{\Gamma(\dsphikkp)} = 0.540\pm 0.033\pm0.048
\ee
Our number is comparable with all the other measurements in this channel,
and the new world average of $\Gamma(\phi\mu\nu)/\Gamma(\phi\pi)$ is
0.540 $\pm$ 0.040. 

\subsection{The Form Factor Ratios of $\dksmn$}
We measured the form factor ratios of $\dksmn$ and it charge conjugate decays
with consideration on the S-wave contribution. Our study shows that the
effect of S-wave on the measurement is minimal while the effect of charm
background is significant. The new FOCUS results are as follows~\cite{focus_ff},
\bea
   & R_V = 1.504 \pm 0.057 \pm 0.039 \\
   & R_2 = 0.875 \pm 0.049 \pm 0.064
\eea 
Our $R_V$ value is 2.9 $\sigma$ below the E791 measurements~\cite{e791},
but consistent with others. Our $R_2$ value is consistent with other
measurements. The new world averages are 1.66 $\pm$ 0.060 and
0.827 $\pm$ 0.055 for $R_V$ and $R_2$, respectively.

\section{Note on the Hadronic Decays of Charm Particles} 

 The proper interpretation of the hadronic decays is more complicated than expected. We observed that
Final State Interactions (FSI) play a central role in the hadronic decays. 
 For example our recent analysis on the branching ratio $\Gamma(D^{0} \rightarrow
 K^-K^+)/\Gamma(D^{0}\rightarrow \pi^-\pi^+)$~\cite{focus_kkpipi}, 
confirm that FSI are fundamental. Actually an isospin analysis of the channels $D\rightarrow KK$ and 
$D\rightarrow \pi \pi $ reveals that the elastic FSI cannot account for all the large deviation from 
unity (we measure $2.81\pm 0.10\pm 0.06$) of this ratio. The most reasonable explanation seems to be 
inelastic FSI that also allow for the transition $KK \rightarrow \pi\pi$.

 For the multibody modes, where resonances are present, we think that the amplitude analysis 
(Dalitz plot analysis) is the correct way to determine the resonant substructure of the decays. 
An extensive program of Dalitz plot analyses is going on for the 3-body final states. Actually FOCUS 
is conducting a pioneer work using, for the first time in the analyses of charm decays, the formalism 
of K-matrix.

 As an example consider the \emph{CP}-odd state $K^0_s \phi$ from the decay mode
$D^0\rightarrow K^0_s K^-K^+$; one cannot get a {\it pure} \emph{CP}-odd eigenstate
near the $\phi(=K^-K^+)$ region because of the presence of the \emph{CP}-even
$K^0_s f_0$ decaying into the same final state. Instead a Dalitz plot analysis is necessary 
to determine properly the relative fractions. And this is valid also for the beauty decay mode
\hbox{$B^0\rightarrow K^0_s K^-K^+$}.
 
\section{Conclusions}
 
 The FOCUS experiment has measured the lifetime of all the weakly decaying singly 
charmed particles, establishing the charm lifetime hierachy.

 We found new S-wave interference phenomena in $\dkpmn$ decays. Considering this 
effect in further analyses, we measured the branching ratio $\Gamma(D^+ \ra K^*\mu\nu)/\Gamma(D^+ \ra K\pi\pi)$ 
and the form factor ratios of $\dkpmn$ decays with improved statistical errors. We also measured 
the branching ratio $\Gamma(D_s \ra \phi\mu\nu)/\Gamma(D_s \ra \phi\pi)$. 

 This lead us to the following question: will there be similar effects (interference) 
in other charm semileptonic or beauty semileptonic channels?
 
 We will see, in the meanwhile the analyses in other semileptonic charm decay modes are actively going 
on and we expect new results soon.
 
 At 30 years from the discovery of the {\it c} quark the physics analyses of the first heavy quark
have reached a complete maturity. With the large statistics now available in the charm sector we start
to see unexpected effects which complicate the interpretation of the decay processes, both in semileptonic
and hadronic decays.

\end{document}